\documentclass[fleqn,usenatbib]{mnras}
\usepackage{newtxtext,newtxmath}
\usepackage[T1]{fontenc}
\usepackage{ae,aecompl}
\usepackage{color} 

\usepackage{graphicx}	
\usepackage{amsmath}	

\title[CP Stars in Integrated Light]{Chemically Peculiar Stars in Integrated Light Stellar Population Models and Local Group Galaxies}

\author[G. Worthey \& X. Shi]{
Guy Worthey$^{1}$\thanks{E-mail: gworthey@wsu.edu} 
Xiang Shi,$^{1}$ 
\\
$^{1}$Washington State University, Pullman, WA 99164, USA\\
}

\date{Accepted 2022 November 04. Received 2022 October 16; in original form 2022 February 09}

\pubyear{2022}

\begin{document}
\label{firstpage}
\pagerange{\pageref{firstpage}--\pageref{lastpage}}
\maketitle

\begin{abstract}
Integrated-light models that incorporate common types of chemically peculiar (CP) stars are assembled using synthetic spectra. Selected spectral features encode significant age information for populations with ages $\sim$50 Myr $<$ age $< \sim$2 Gyr. Due to the alleviation of template mismatch, the inclusion of CP star features in model spectra improves the accuracy of recovered stellar population parameters, but we are not able to show that new or unique age information can be extracted from the weak CP features compared to continuum fitting and strong-feature strengths, at least at the present state of the art. An age-extraction routine that recovers 2- and 3-burst age structures is employed to analyze the spectra of local group galaxies. NGC~224 (M~31) has a stellar population too old for the types of CP stars we examine. NGC~221 (M~32) also shows no CP spectral features. It appears to contain a component at age $\sim$1 Gyr at 1\% by mass in addition to its dominant 4.7 Gyr population. Unlike SDSS galaxy spectrum averages, NGC~205 (M~110) contains no features due to HgMn stars. This excludes the age range associated with HgMn production, and its near-nuclear spectrum is best fit by a 68$\pm$2 Myr population superimposed on an older population with a 1.85$\pm$0.1 Gyr component. Both NGC~205 and NGC~221 have an ancient component whose mass is not easy to constrain given the overwhelming light-dominance of the younger populations.
\end{abstract}


\section{Introduction}

The formation history of galaxies continues to be a topic of keen interest, unsolved  because of its complex nature in which star formation, chemical evolution, and dynamical evolution are intertwined. Progress in one of these areas will help the others, and this paper concerns the question of unraveling star formation history from the integrated starlight of galaxies. Integrated light age indicators include ultraviolet colours \citep{1996AJ....112..948W,1997AIPC..408..175D} and Balmer feature absorption strengths \citep{1994ApJS...95..107W,2000AJ....119.1645T}, when these indicators are arrayed against metal-sensitive features such as metallic absorption lines. A light-weighted mean age and mean metallicity can be derived for galaxies without nebular emission using Balmer versus metal line pairs. This relatively blunt instrument reveals a mean age and mean heavy element abundance, but cannot reveal a star formation history. Furthermore, significant uncertainties remain because of scatter in the various integrated-light models \citep{2021A&A...651A..99P}.

In an attempt to accelerate the quest for believable multi-bin star formation histories, \cite{2015A&A...580L...5W} suggested that common types of chemically peculiar (CP) stars could be used to better disentangle an age history from integrated light spectroscopy. The relevant stars are those with radiative atmospheres on and near the main sequence. They frequently show patterns of strong elemental enhancement in their spectra \citep{1996Ap&SS.237...77S,1993ASPC...44..577N} due to radiative levitation \citep{1970ApJ...160..641M}. The fraction of stars that exhibit this signature is high enough such that the spectral signature survives to appear in integrated light at subtle levels. Furthermore, the character of the spectral signature changes with main sequence temperature. In stellar populations, the main sequence is the primary age indicator. As the main sequence turnoff cools, the various CP types should contribute strongly, and then disappear. This on-off behavior is even more nonlinear than Balmer line strength, and could in principle provide a much more secure multi-bin age indicator.

In this work we consider four classes of the most common CP stars. 
The Ap stars represent the strongly-magnetic branch of the CP stars \citep{1996Ap&SS.237...77S} and span spectral types B6-F4. Ap stars typically show overabundances of silicon, strontium, chromium and europium. In addition, large overabundances are often seen in praseodymium and neodymium. 
An Am star or metallic-line star is a type of chemically peculiar star of spectral type A or early F whose spectrum has strong absorption lines of metals such as zinc, strontium, zirconium, and barium, and deficiencies of others, such as calcium and scandium \citep{Conti_1970}, albeit with considerable variation from star to star.
Mercury-manganese (HgMn) stars span spectral classes B7 through B9 and temperatures from 10700 to 14000 K \citet{2009ssc..book.....G}. Indicative features include Hg3984, Mn3944, Mn4137,
Mn4206, and Mn4282. Along with Am stars, Hg-Mn stars seem to represent a weakly-magnetic sequence.
The hottest stars we consider here are helium-weak stars, deficient in He compared to normal stars \citep{bohlenderlandstreet1988}. He-weak stars span spectral types B2-B8. We neglect hotter He-strong and Wolf-Rayet stars for this paper, though of course they should be considered fully in future work.

The ability of these CP stars to provide integrated light age information is clear in principle, but detectability is an issue. \cite{2015A&A...580L...5W} detected HgMn stars in averaged Sloan Digital Sky Survey (SDSS) galaxies \citep{2012MNRAS.420.1217D}, but cautioned that high signal observations would be required to fully exploit the washed out absorption lines of interest. In this paper we develop models for integrated light that include the CP stars, then apply them to single galaxies, the near-nuclear regions of local group galaxies NGC~205, NGC~221, and NGC~224.

In $\S$2, we describe new integrated light models that schematically include synthetic CP star spectra. We perform tests with the models, including model inversions in $\S$3 in order the gauge the benefit of including CP star information. We present observations of local group galaxy nuclei in $\S$4. We discuss the results, including signal detectability and the promise of eventual success of this technique in $\S$5.

\section{Models}

The integrated light models begin within the framework of \cite{1994ApJS...95..107W} models, They incorporate Padova evolution \citep{2008A&A...482..883M} with subsequent updates. Predictions for observables include a set of empirically-zeroed absorption feature indices \citep{2017MNRAS.467..674T} and synthetic spectra. 

A suite of new synthetic spectra was calculated to accommodate the need to include representative CP star species using ATLAS9 and SYNTHE \citep{1993ASPC...44...87K,1993KurCD..18.....K}.  A recipe for how the surface abundances should develop as a function of rotation, age, spotting, and magnetic field strength would be desirable, but such a recipe does not exist. As a simplified first attempt, we assume that each temperature bin along the main sequence contains a constant fraction of CP stars. 

For this work, a median abundance pattern for each CP type is adopted from literature sources. Ultimately, this is not the proper way to weight their contributions, because the surface abundance patterns of CP stars show large astrophysical scatter. We also average over closely related spectral types. For example, Ap stars of the Si, SiCrEu, SrCrEu and Sr classes are lumped into one category. As for literature sources, for Am stars, most of the abundances were taken from \cite{2014PASP..126..345Y} with reference to \cite{2018MNRAS.480.2953G}. Data for other CP types were collected from studies of individual stars or small groupings of stars \citep{2003A&A...402..331A,2014A&A...561A.147B,2008MNRAS.391..901F,2007MNRAS.376..361F,1986ApJ...304..425T,2007A&A...468..263C,1989MNRAS.239..487A,2001A&A...367..597A,1994MNRAS.266...97A,1988MNRAS.235..763A,1998MNRAS.297....1A,1998MNRAS.300..359A,2006A&A...447..685A,1988MNRAS.230..671A,1992MNRAS.258..167A,1996MNRAS.283.1115R,1997MNRAS.288..470A,1999MNRAS.305..591A,1999MNRAS.310..146A,2011AN....332..681Y,1994MNRAS.271..355A,2011AN....332..153A,2010AN....331..378A,1997MNRAS.288..501C}. Several elements are well studied and secure, but data for others is sparse. Our medians are summarized in Table \ref{tab:mix}, where the number of significant digits encodes uncertainty. For some elements, only order-of-magnitude estimates are available, guided by the general pattern of overabundance by species mass. These are indicated by colons in Table \ref{tab:mix}. 

The abundance of at least one element, Ti, trends strongly with temperature among Ap stars. Similarly, \cite{2018MNRAS.480.2953G} points out a temperature-dependent Ca abundance trend among Am stars. We ignore these for the current model generation. A general rule is that, even when abundances seem to trend smoothly, the interplay of excitation and ionization in the stellar atmospheres can make the observed absorption line strengths vary much more strongly \citep{1925PhDT.........1P}. 

\begin{table}
  \caption{Observed abundance patterns, given as [X/H].}
  \label{tab:mix}
    \centering
    \begin{tabular}{r|c|c|c|c|c}
    \hline
$Z$ & elem. & Am  & Ap    & Hg-Mn  & He-wk \\
    \hline
 2 & He & $-0.2$ & $-1.3$ & $-0.8$ & $-0.5$ \\
 6 & C  & $0.0$  & $-1.0$ & $0.0$  & $0.0$  \\
 8 & O  & $0.0$  & $-0.4$ & $0.0$  & $-0.4$ \\
10 & Ne & \ldots & $0.4$  & \ldots & $-0.2$ \\
11 & Na &$-1.5$: &$-0.25$ & \ldots & \ldots \\
12 & Mg & $-0.2$ & $-1.1$ & \ldots &$-0.25$ \\
13 & Al & $0.3$  & $-0.5$ & $-1$   & $0$    \\
14 & Si & $-0.10$& $0.85$ & $-0.10$& $0.13$ \\
15 & P  & \ldots & \ldots & $1.5$  & $0.8$  \\
16 & S  & $0.6$  & $0.0$  & $-0.2$ & $-0.4$ \\
20 & Ca & $-0.3$ & $0.2$  & $0.4$  & $0.3$  \\
21 & Sc & $-0.5$ & $0.0$  & $0.7$  & $1.4$  \\
22 & Ti & $0.2$  & $1.1$  & $0.7$  & $1.4$  \\
23 & V  & $0.6$  & $0.0$  & $0.0$  & \ldots \\
24 & Cr & $0.2$  & $1.7$  & $0.4$  & $1.0$  \\
25 & Mn & $0.2$  & $0.5$  & $2.0$  & $1.9$  \\
26 & Fe & $0.3$  & $0.6$  & $0.1$  & $0.9$  \\
27 & Co & $0.6$  & $1.2$  & \ldots & \ldots \\
28 & Ni & $0.6$  & $0.0$  & $-0.5$ & $0.3$  \\
29 & Cu & \ldots & $-0.4$ & \ldots & $1.3$  \\
30 & Zn & $0.7$  & $-0.7$ & $1.0$  & \ldots \\
31 & Ga & \ldots & \ldots & $3.5$  & \ldots \\
38 & Sr & $0.8$  & $1.0$  & $2.0$  & $2.1$  \\
39 & Y  & $1$:   & $1.3$  & $2.5$  & $3$    \\
40 & Zr & $1$    & $0.4$  & $2$    & $3$    \\
54 & Xe & \ldots & \ldots & $4$    & $4$    \\
56 & Ba & $1.5$  & $1.75$ & $0.8$  & \ldots \\
57 & La & $1.2$  & $2.0$  & $1.5$  & \ldots \\
58 & Ce & \ldots & $3.5$  & $3.0$  & $4.1$  \\
59 & Pr & \ldots & $3.6$  & $3.3$  & $5.3$  \\
60 & Nd & \ldots & $3.3$  & $2.0$  & $4.5$  \\
62 & Sm & \ldots & $2.2$  & \ldots & $5$    \\
66 & Dy & \ldots & $2.7$  & \ldots & $4.5$  \\
80 & Hg & \ldots & $3.2$  & $5$    & $4.5$  \\
    \hline
    \end{tabular}
\end{table}

\begin{table}
  \caption{CP star number fractions.}
  \label{tab:fracs}
    \centering
    \begin{tabular}{r|c|l|l|l|l}
    \hline
    $T_{\rm eff}$ & Sp. &  Am  &    Ap    &  Hg-Mn & He-wk \\
    \hline
 6650  & F4 & 0.01   & \ldots & \ldots  & \ldots \\
 6760  & F3 & 0.10   & 0.02    & \ldots  & \ldots \\
 7000  & F2 & 0.08   & 0.04    & \ldots  & \ldots \\
 7120  & F1 & 0.03   & 0.00    & \ldots  & \ldots \\
 7250  & F0 & 0.22   & 0.03    & \ldots  & \ldots \\
 7380  & A9 & 0.12   & 0.00    & \ldots  & \ldots \\
 7830  & A7 & 0.30   & 0.03    & \ldots  & \ldots \\
 7990  & A6 & 0.45   & 0.15    & \ldots  & \ldots \\
 8150  & A5 & 0.35   & 0.02    & \ldots  & \ldots \\
 8335  & A4 & 0.40   & 0.02    & \ldots  & \ldots \\
 8520  & A3 & 0.24   & 0.02    & \ldots  & \ldots \\
 8900  & A2 & 0.22   & 0.07    & \ldots  & \ldots \\
 9500  & A1 & 0.08   & 0.10    & \ldots  & \ldots \\
 9800  & A0 & 0.05   & 0.18    & \ldots  & \ldots \\
10700  & B9 & \ldots& 0.19    & 0.12     & 0.01    \\
11500  & B8 & \ldots& 0.05    & 0.13     & 0.06    \\
13000  & B7 & \ldots& 0.02    & 0.07     & 0.06    \\
14000  & B6 & \ldots& 0.05    & \ldots  & 0.04    \\
15000  & B5 & \ldots& 0.00    & \ldots  & 0.06    \\
15750  & B4 & \ldots& \ldots & \ldots  & 0.06    \\
16500  & B3 & \ldots& \ldots & \ldots  & 0.05    \\
19500  & B2 & \ldots& \ldots & \ldots  & 0.03    \\
24500  & B1 & \ldots& \ldots & \ldots  & 0.01    \\
    \hline
    \end{tabular}
\end{table}

We computed new sets of LTE stellar atmospheres and emergent spectra with R. L. Kurucz's ATLAS9 and SYNTHE programs, all at the solar mixture, and then the solar mixture with the Table \ref{tab:mix} changes applied. Microturbulence was held constant at 2 km s$^{-1}$ and no rotational convolutions were performed. The spectral changes were applied differentially within the integrated light models to accommodate mild departures from the base (solar) abundance level, but these models should not be applied to inherently metal poor systems such as dwarf galaxies.

\begin{figure}
	\includegraphics[width=\columnwidth]{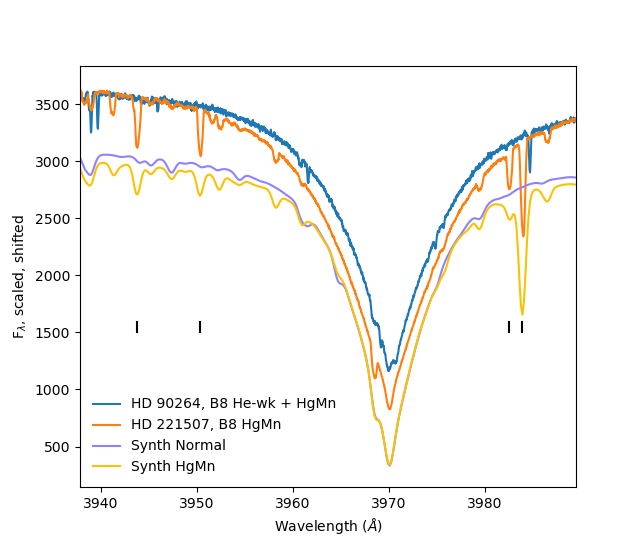} 
    \caption{Spectra of HgMn star HD221507 (orange) and spectroscopic binary HD90264 whose dominant component is a He-weak star of the same spectral type (blue). Two synthetic spectra at $T_{\rm eff}=13000$ K and log $g$ = 3.5 are shown at solar ratios (purple) and with a HgMn abundance pattern (yellow), shifted downward for clarity. Vertical lines mark some stronger diagnostic absorption lines due to Mn and Hg (black vertical dashes). }
    \label{fig:HgMn_4750}
\end{figure}

In order to assess the fidelity of the synthetic spectra, graphical comparisons were made with CP and normal stars in the UVES-POPS spectral library \citep{2003Msngr.114...10B}. For example, in Figure \ref{fig:HgMn_4750}, B8 HgMn star HD221507 is compared to HD90264, a double-lined spectroscopic binary with a He-weak primary and HgMn secondary (the latter being redshifted in the figure) \citep{2010A&A...521A..75Q} and roughly the same spectral type.
Synthetic spectra are also shown, one at solar abundance ratios and one with an HgMn abundance pattern.
With an exception or two, the line list appears to match the wavelengths and strengths of the lines. This statement holds true for all wavelengths and all CP types except Am, for which the observed selection of stars proved sparse.

\begin{figure}
	\includegraphics[width=\columnwidth]{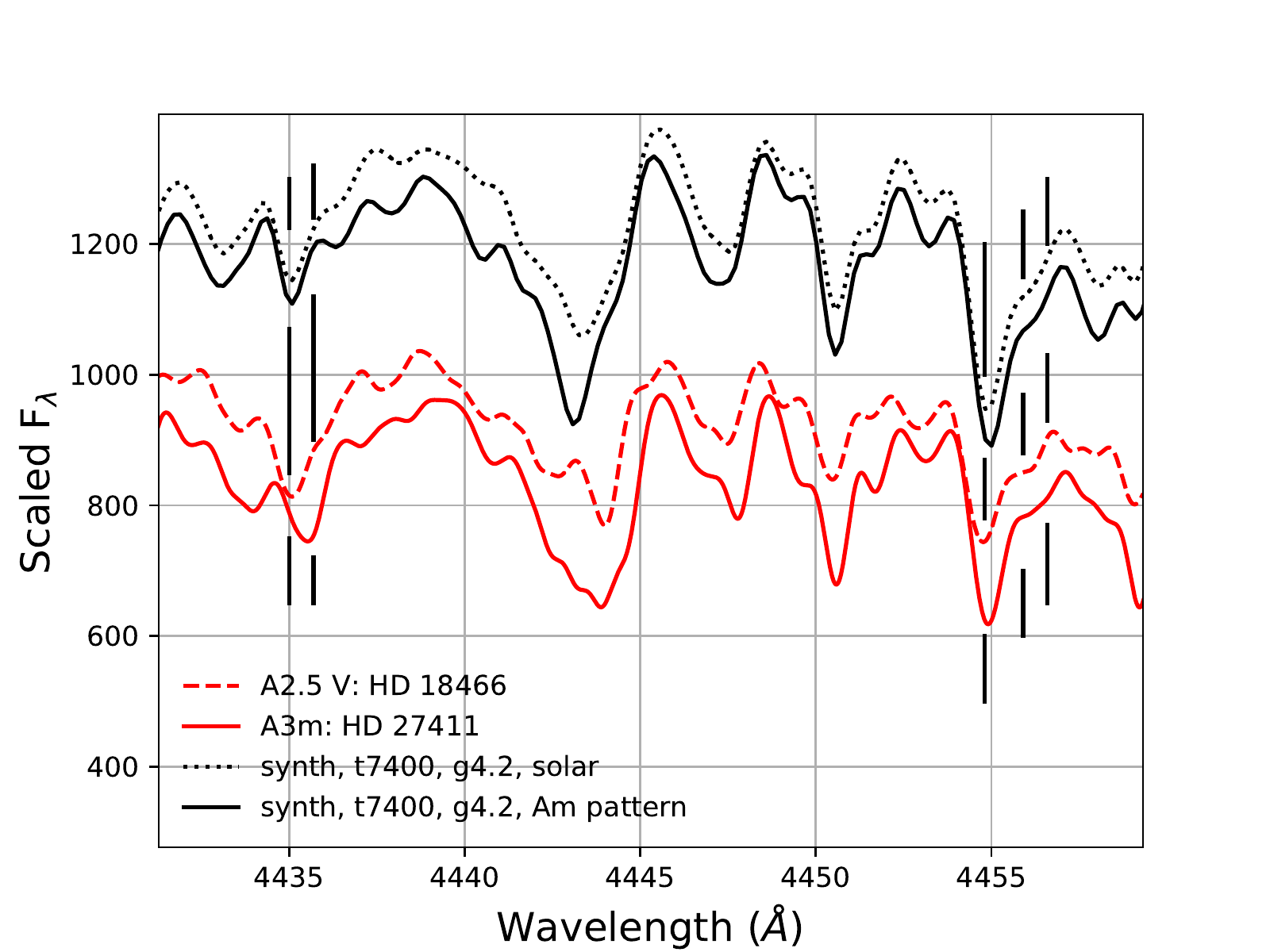}  
    \caption{Spectra that include calcium lines at $\lambda\lambda$4435.0, $\lambda\lambda$4435.7, $\lambda\lambda$4454.8, $\lambda\lambda$4455.9, and $\lambda\lambda$4456.6 are shown for A2.5 V star HD~18466 (red dashed) and A3m star HD~27411 (red solid), along with synthetic spectra at $T_{\rm eff} = 7400$ and log $g$ = 4.2, one with normal abundance (black dotted) and one with the Am abundance pattern (black solid). The synthetic spectra were scaled to match the observations, then shifted by 250 units for purposes of display. Although Ca is underabundant in the Am mixture ([Ca/H] $= -0.3$), due to competition with other species that become more abundant rather than less, the position of the line on the curve of growth, and overall continuum depression, the Ca lines do not weaken in the synthetic Am spectrum except for $\lambda\lambda$4435.7. Some of the line-strengthening in HD~27411 could be from a cooler temperature: the two stars are not perfect analogs.}
    \label{fig:am1}
\end{figure}

In Fig. \ref{fig:am1}, Am star HD~27411 is compared to normal A2.5 V star HD~18466, and also to synthetic spectra at $T_{\rm eff} = 7400$ K and log $g$ = 4.2, one with the solar mixture, and one with the Am mixture from Table \ref{tab:mix}. Data is from UVES-POPS \citep{2003Msngr.114...10B}. One signature of Am stars is a weakening of Ca lines, but the figure shows none. We assumed [Ca/H] = $-0.3$ for the synthetic Am abundance pattern, and \cite{2012MNRAS.421.1222C} derive [Ca/H] = $-0.3$ for HD~27411, but in both cases the Ca absorption strengthens. This is presumably due to the fact that, when metals are added, the line blanketing and extra opacity due to free electrons depresses the continuum. The continuum depression make measuring the Am effect more difficult than it would otherwise have been. Because a weakening of Ca and Sc is a signature of Am stars in general, we may have been too conservative with our mild [Ca/H] = $-0.3$ assumption, and it should be more extreme to match the average Am abundance pattern. Overall, the synthetic reproduction of the observational spectrum is encouragingly good. Bolstered by these comparisons and others, we computed integrated-light models with the addition of the CP phenomenon.

During the integration along the isochrone as the models are being calculated, the new synthetic CP spectra are blended with normal-mixture spectra according to the proportions of \cite{1996Ap&SS.237...77S} as a function of spectral type with conversions to $T_{\rm eff}$ from \citet{2009ssc..book.....G}. This recipe is summarized in Table \ref{tab:fracs}.

\begin{figure}
	\includegraphics[width=\columnwidth]{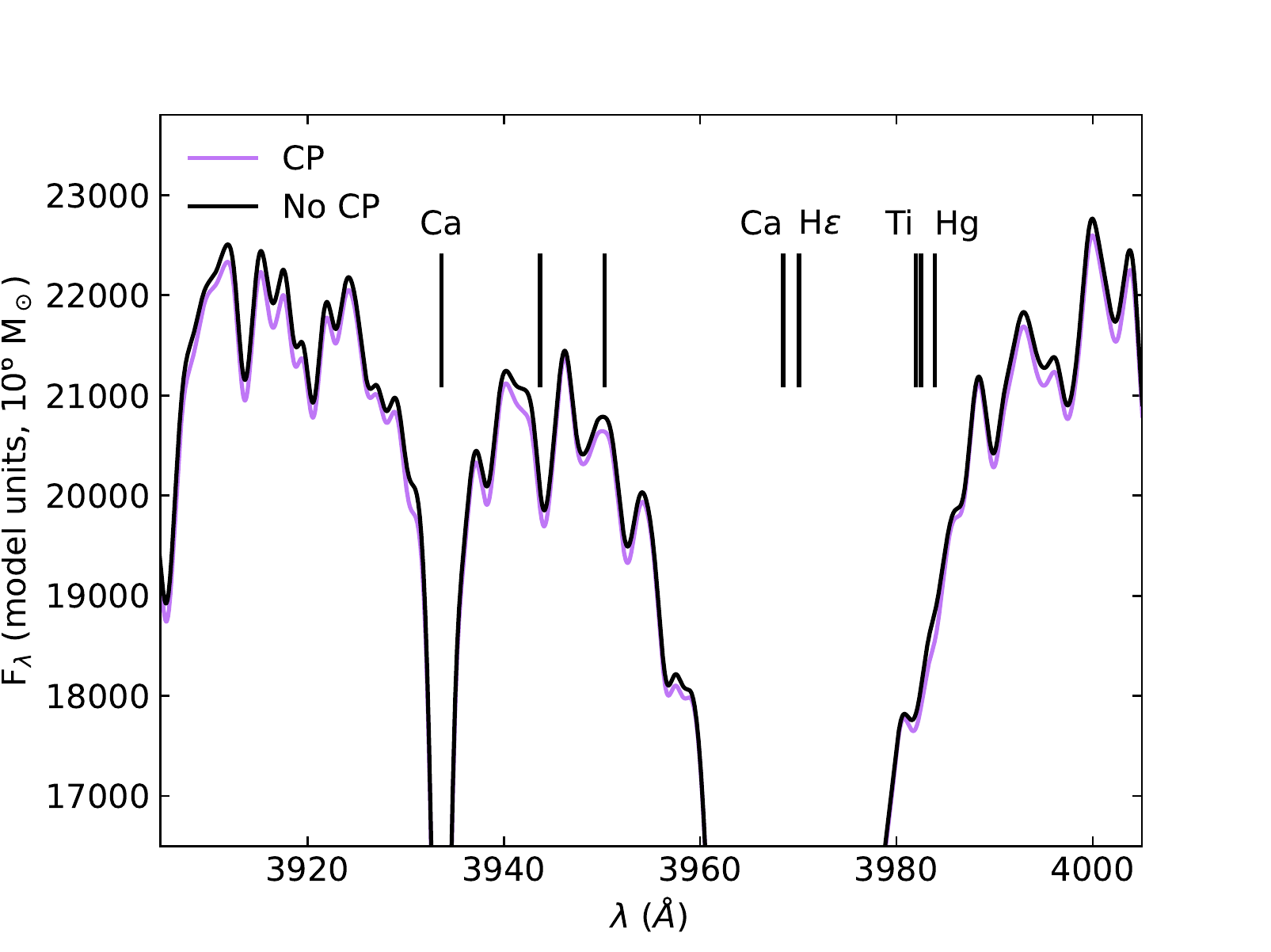} 
    \caption{A simple stellar population, age 125 Myr, solar composition, smoothed to 50 km s$^{-1}$ velocity dispersion, with (purple) and without (black) CP stars mixed in. A few lines due to Mn (unmarked) and other species (marked) are indicated. The features used for CP stellar classification do not necessarily translate to the largest effects in integrated light.}
    \label{fig:125Myr}
\end{figure}

Then, with models in hand, we step toward assessing whether they will be useful in the analysis of galaxy spectra. Step one is to discover spectral features that change when CP stars are added to the mix. Figure \ref{fig:125Myr} shows a spot-check for a 125 Myr simple stellar population at solar metallicity and rendered at 50 km s$^{-1}$. An overall continuum depression is accompanied by increased absorption at the wavelengths most influenced by the CP-altered chemical composition. 

\begin{figure*}
	\includegraphics[width=\textwidth]{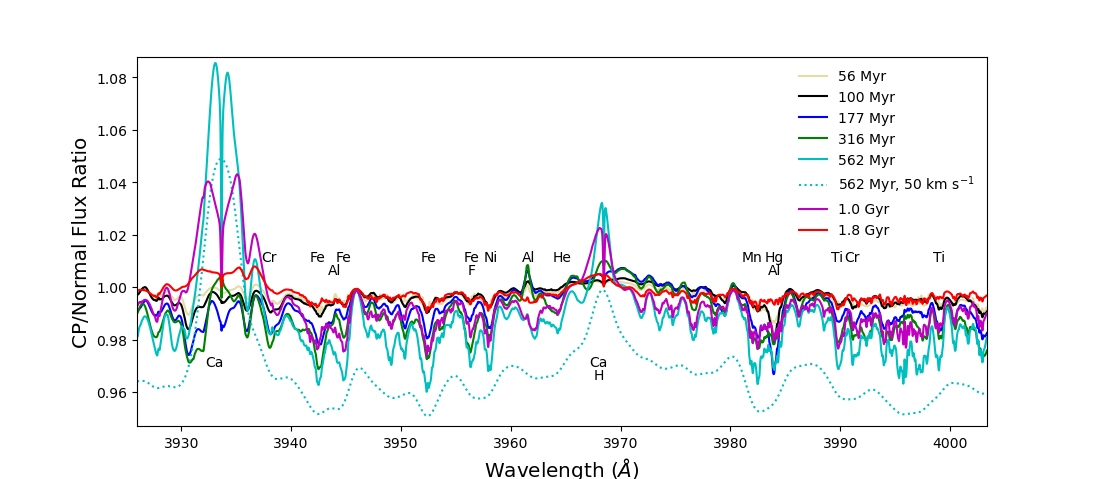} 
    \caption{High-resolution SSP model spectra near Ca H\&K, SSPs with CP stars included divided by SSP model spectra without. Ages are shown according to the caption, and the SSPs are at solar abundance. A few contributing species are marked. A velocity-smoothed version of the 562 Myr model is shown for reference, displaced downward by 0.02 for clarity. Am stars switch to HgMn stars between the 562 Myr and 316 Myr examples.}
    \label{fig:Ca3933}
\end{figure*}

\begin{figure}
	\includegraphics[width=\columnwidth]{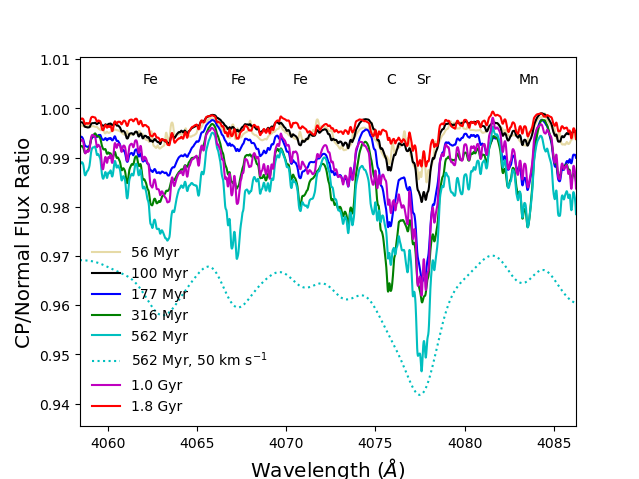} 
    \caption{High-resolution SSP model spectra in the region of Sr4077, SSPs with CP stars included divided by SSP model spectra without. Line types as in Fig. \ref{fig:Ca3933}. }
    \label{fig:Sr4077}
\end{figure}

\begin{figure}
	\includegraphics[width=\columnwidth]{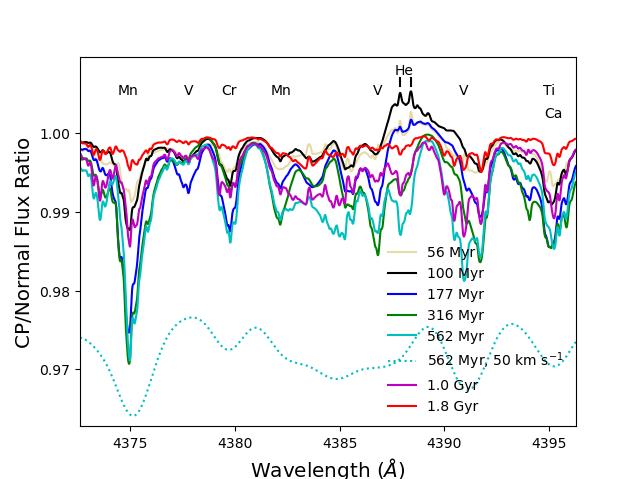}  
    \caption{Ratios of SSP model spectra in the region of He4387 are shown at high resolution, CP stars included divided by CP stars excluded. Line types as in Fig. \ref{fig:Ca3933}. The effects of He-weak stars appear at the youngest ages.}
    \label{fig:He4387}
\end{figure}

Spectral variations with population age can be visualized by a division of CP/normal model spectra, as in Figs. \ref{fig:Ca3933}, \ref{fig:Sr4077}, and \ref{fig:He4387}. At high spectral resolution, the effects at $\lambda < 4000$\AA\ are $\sim$5\%, while in the red, the effects are of order 1\%. These amplitudes will diminish if a target galaxy has a significant velocity dispersion, and we give a visual impression in Figs. \ref{fig:Ca3933}, \ref{fig:Sr4077}, and \ref{fig:He4387}, with a trace smoothed to 50 km s$^{-1}$. 

One must also keep in mind, especially in the violet portions of the spectrum, that atomic absorption lines are extremely crowded. Any particular absorption feature is likely to be a blend of several. $\lambda\lambda$3944, for example, is mostly Al for normal stars, but the Al weakens and Mn dominates for HgMn stars. Sr II 4077 is blended with Si II and Cr II. The Si II double line at 4128, 4131 can be blended with and sometimes dominated by Eu II, but Fe I, Fe II, Ce II, and Gd II can also be present \citep{2009ssc..book.....G}. Each atomic transition waxes and wanes with temperature and therefore SSP age. The feature labels in Figs. \ref{fig:Ca3933}, \ref{fig:Sr4077}, and \ref{fig:He4387} indicate some contributor to absorption, but not necessarily the dominant one for all ages. One also must acknowledge the approximate and incomplete nature of the line-blanketed synthetic spectra upon which the SSP models depend. As amazing as they are, they never exactly reproduce every observed spectral feature.

The division of CP/normal models such as in Figs. \ref{fig:Ca3933}, \ref{fig:Sr4077}, and \ref{fig:He4387} allows us to spot advantageous wavelengths at which to measure the effect. However, it gives no information on whether including CP stars will make finding galaxy ages more secure. To explore that, we selected some dozens of pairs of nearby wavelengths and formed flux ratios. The wavelengths typically sampled absorption bottoms and paired them with nearby pseudocontinua. We then examined the behavior of these ratios with age. 

\begin{figure}
	\includegraphics[width=\columnwidth]{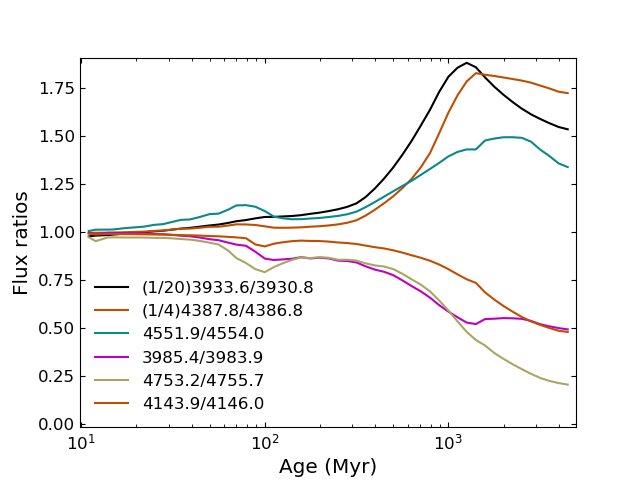}  
    \caption{Ratios of flux point pairs are shown as a function of SSP age. The idea here is that, given a set of such flux ratios, they would map to age uniquely. Many of them rise and fall as CP star flavors shift, in contrast to, for example, broad-band colours, which would monotonically increase with age. These ratios were constructed from high resolution spectra, and thus would decrease in range in the case of significant velocity dispersion.}
    \label{fig:fluxratio}
\end{figure}

A few examples are are shown in Fig. \ref{fig:fluxratio}. The ones shown are a good sampling of the various behaviors exhibited by the ratios (we tested dozens). Distinct signatures show up at $\sim 100$ Myr and $\sim 1$ Gyr, where the ebb and flow of CP stars introduce altered spectra. In addition, these observable spectral trends do not exist by themselves. Including additional quantities such as Balmer line indices or continuum tracers such as photometric colours can only improve age discrimination. 

\section{A model inversion test}

Figure \ref{fig:fluxratio} serves to show that, at high spectral resolution, the addition of CP stars fulfills its promise to provide additional age-discriminatory information extractable from integrated galaxy spectra. In practice, this age information is fairly well buried in the spectrum. To extract the age information, some statistical method will be employed to consider simultaneously overall metallicity, metallicity spread, abundance ratios, initial mass function, and age effects \citep{2012MNRAS.421.1908J,2014ApJ...783...20W,2014ApJ...780...33C}.

For this work, we step toward that (rather daunting) list of effects by staying in the domain of simple stellar populations (SSPs, bursts of a single age and metallicity). We furthermore restrict our exploration to solar metallicity, and we do not attempt to include empirical stellar libraries, but stay in the synthetic domain. These interim models will hopefully point out the path for future development of the method.

\begin{figure}
	\includegraphics[width=\columnwidth]{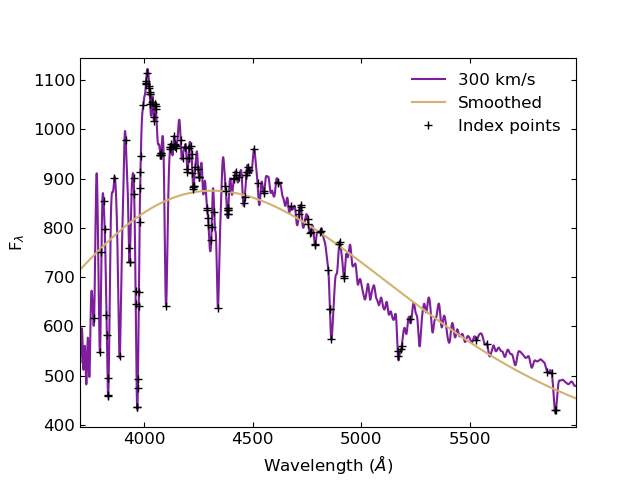}  
    \caption{200 Myr, solar SSP based on synthetic spectra with the average continuum (smooth line) and selected wavelengths that sample hopefully-diagnostic heavy metal lines, H lines, and He lines (black plusses). Flux/continuum ratios at the selected wavelengths become indices. The SSP spectrum has been smoothed to a velocity dispersion of $\sigma = 300$ km s$^{-1}$ for clarity.}
    \label{fig:setup}
\end{figure}

To approach the inversion problem, that is, to discover the population parameters from the observed spectrum, we wished to exclude (if possible) continuum-fitting. The CP method must rely on specific spectral features. An inversion program from which one expected astrophysics to emerge would consider the shape of the continuum, along with any other constraints that are available. But we limit our scope to the CP technique in isolation, as much as possible. This is probably an impossible request, but our attempt is described as follows and illustrated in Figure \ref{fig:setup}. A 10-term polynomial fit to the spectra, which range over 3000\AA\ $< \lambda <$ 10000\AA, forms a reference flux level. It isn't a continuum in any sense, but it serves to anchor the flux points and we sometimes call it the average continuum. We then select 162 wavelengths at which we expect interesting behavior. Most often, these are the cores of the lines of species we wish to track, but some are nearby (actual) pseudocontinuum points that show diagnostic behavior, isolated by means of figures like Figs. \ref{fig:Sr4077} and \ref{fig:He4387}. The wavelengths we track are shown in Fig. \ref{fig:setup} and listed in Table \ref{tab:specks}. The set of observables we model are flux/continuum ratios at each of the Table \ref{tab:specks} wavelengths.

\begin{table}
  \caption{Index wavelengths.}
  \label{tab:specks}
    \centering
    \begin{tabular}{l|l|l|l|l}
    \hline
    $\lambda$ (\AA) & $\lambda$ (\AA) &  $\lambda$ (\AA) &    $\lambda$ (\AA)   &  $\lambda$ (\AA)  \\
    \hline
H       & 3832.2 & 4079.0 & 4376.7 & 4812.2  \\
3770.633& 3863.6 & 4120.9 & 4386.8 & 4848.2  \\
3797.909& 3916.0 & 4123.4 & 4387.8 & 4855.4  \\
3835.397& 3930.8 & 4139.2 & 4388.4 & 4897.7  \\
3889.064& 3933.6 & 4142.8 & 4414.2 & 4899.9  \\
3970.075& 3952.3 & 4146.0 & 4415.6 & 4920.6  \\
4101.734& 3954.8 & 4168.9 & 4422.6 & 4922.3  \\
4340.472& 3961.5 & 4177.5 & 4425.4 & 5167.3  \\
4861.35 & 3962.2 & 4186.5 & 4437.9 & 5168.7  \\
6562.79 & 3968.3 & 4187.0 & 4461.6 & 5183.5  \\
He      & 3971.9 & 4198.5 & 4463.8 & 5185.9  \\
3834.22 & 3972.6 & 4200.1 & 4470.8 & 5222.1  \\
4009.71 & 3975.8 & 4205.0 & 4471.7 & 5526.8  \\
4025.14 & 3976.6 & 4207.7 & 4478.3 & 5578.1  \\
4026.17 & 3980.2 & 4215.6 & 4481.2 & 5853.6  \\
4120.2  & 3982.5 & 4218.8 & 4505.3 & 5875.7  \\
4121.3  & 3983.9 & 4222.5 & 4522.5 & 5889.9  \\
4143.4  & 3985.4 & 4225.4 & 4551.9 & 5895.7  \\
4143.86 & 3997.2 & 4227.6 & 4554.0 & 6115.0  \\
4387.89 & 4008.9 & 4228.9 & 4616.4 & 6122.2  \\
4388.4  & 4009.1 & 4236.0 & 4617.7 & 6126.2  \\
4470.86 & 4012.4 & 4249.4 & 4676.8 & 6141.7  \\
Metal   & 4024.6 & 4253.0 & 4713.0 & 6496.9  \\
3494.7  & 4026.4 & 4254.4 & 4714.9 & 6687.1  \\
3495.9  & 4028.4 & 4255.1 & 4720.9 & 6743.1  \\
3497.7  & 4036.5 & 4289.9 & 4722.2 & 6748.7  \\
3513.8  & 4045.5 & 4291.2 & 4753.2 & 6756.9  \\
3515.6  & 4047.8 & 4294.8 & 4755.7 & 8149.3  \\
3804.5  & 4052.5 & 4296.6 & 4764.5 & 8183.2  \\
3814.1  & 4053.8 & 4309.5 & 4767.3 & 8189.3  \\
3819.6  & 4055.0 & 4311.5 & 4786.6 & 8194.8  \\
3827.5  & 4075.8 & 4322.8 & 4788.0 & 8662.0  \\
3828.8  & 4077.6 & 4374.9 & 4810.5 & 8848.0  \\
    \hline
    \end{tabular}
\end{table}

We programmed an inversion program to search within the model grid that predicts the 162 flux ratios. The "observation" in this case was also a model, albeit of known parameters. We tested a two-age model with three parameters (the two ages and the mass ratio between them) and a three-age model with five parameters (the three ages and two mass ratios relative to the third population). When all three age components contributed a fair amount of light to the final "observation" spectrum, the inversion program succeeded in recovering the input ages and mass ratio(s).

\begin{figure}
	\includegraphics[width=\columnwidth]{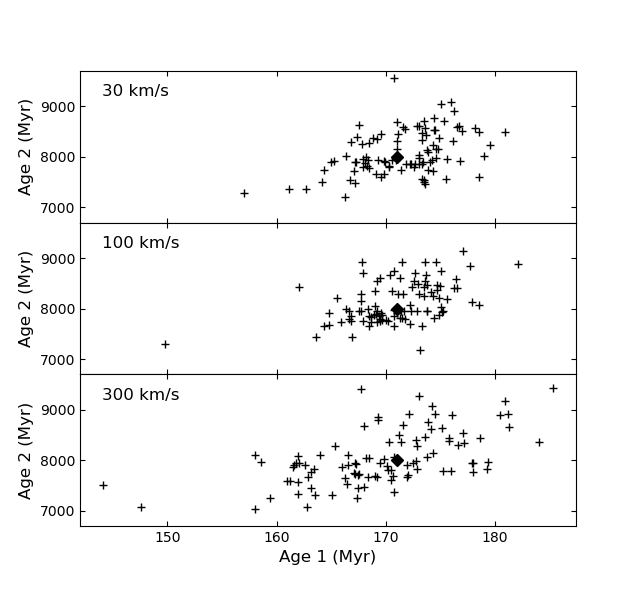}  
    \caption{One hundred two-age Monte Carlo solutions at three velocity dispersions and an assumed S/N of 100. The unperturbed solution is of age 171 Myr added to an 8 Gyr population with mass ratio 0.03 (diamonds). The age scatter is modestly larger for the highest velocity dispersion.}
    \label{fig:MC2}
\end{figure}

While this is good news, it is not yet shown that the success is due to the inclusion of CP features. In fact, we cannot show it. What we show instead is that we have not yet isolated the CP phenomenon well enough to use it effectively as an age-measuring tool. The first clue is that the CP features are clear at high resolution, but fade rapidly for velocity dispersions $\sigma > 30$ km~s$^{-1}$. This behavior should propagate, so that at fixed S/N, the scatter in ages should increase quickly for increased velocity dispersion. Figure \ref{fig:MC2} shows some increase in age scatter, shown in that figure with a 2-age inversion and 100 Monte Carlo trials at S/N = 100, but the increase in age scatter is modest.

The reason that we have fallen short as regards isolating the CP signature is that it is difficult to remove overall flux. Some of our wavelengths of interest are the bottoms of strong lines, such as Balmer lines. For these features, the flux/continuum ratio is mostly measuring continuum change, and so some de-facto continuum-fitting is slipping into the mixture. These strong features are also insensitive to velocity dispersion.

To confirm that we need to refine our approach, we ran sequences of inversions with a target ``galaxy'' composed of one burst at 8 Gyr and one burst at a variable age between 80 and 3000 Myr. Models without the CP enhancement are available to us, so we construct both target and model grid without the improvement (non-CP). We construct targets and model grids with the CP improvement (CP). Finally, we check the effects of template-mismatch by constructing a target ``galaxy'' out of CP models, but using the non-CP grid to attempt parameter recovery. This template mismatch is still very mild, as we are using all other ingredients in exactly the same proportions, and only fractions of stars according to Table \ref{tab:fracs} are swapped in.

\begin{figure}
	\includegraphics[width=\columnwidth]{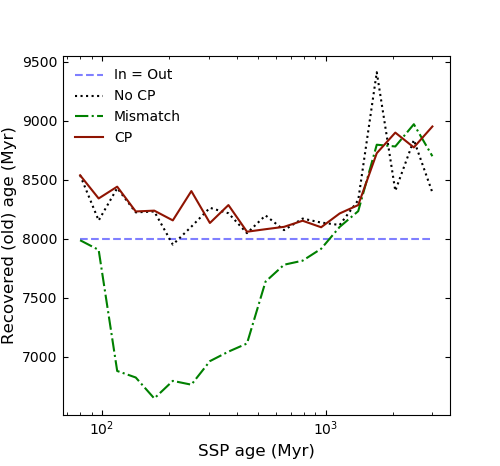}   
    \caption{Summary of recovered (8 Gyr) ages from averages of 100 Monte Carlo trials at S/N = 100 for mass fraction $m_{\rm young}/m_{\rm old} = 0.12$ and a variety of ages for the younger population ($x$ axis). These inversions were performed at $\sigma=30$ km~s$^{-1}$. We show runs with CP target + CP models (red solid), non-CP target + non-CP models (black dotted), and the mixed case of CP target + non-CP models (green dash-dot). Solutions with age$_1 >$ 1000 Myr become unreliable because the luminosity of the younger subpopulation decreases with age to blend in with the older subpopulation. There is a slight bias toward old recovered ages for CP+CP and nonCP+nonCP target-model pairs, and a strong bias toward younger age for the mismatched template case.}
    \label{fig:Q8000}
\end{figure}

\begin{figure}
	\includegraphics[width=\columnwidth]{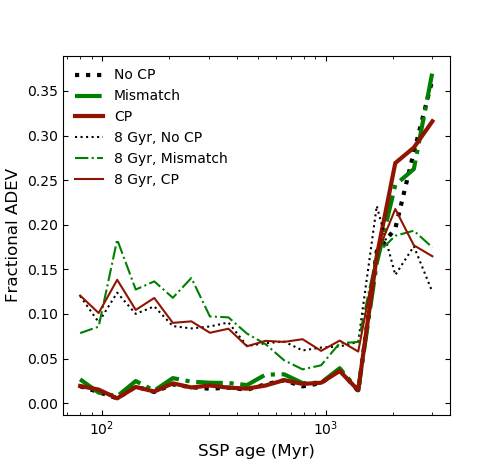} 
    \caption{Monte Carlo scatter in recovered ages, measured as an absolute deviation statistic, for mass fraction $m_{\rm young}/m_{\rm old} = 0.12$ and a variety of ages for the younger population ($x$ axis) and $\sigma = 30$ km~s$^{-1}$. Thicker lines track the younger variable age, while thinner ones track the 8 Gyr component. We show runs with CP target + CP models (red solid), non-CP target + non-CP models (black dotted), and the mixed case of CP target + non-CP models (green dash-dot). Solutions at 1000 Myr and older are unreliable at the assumed S/N (100). As the young component becomes older and dimmer, its ADEV scatter increases while the old-component ADEV decreases.}
    \label{fig:QADEV}
\end{figure}

The results of this exercise are summarized in Figs. \ref{fig:Q8000} and \ref{fig:QADEV} for 2-burst solutions with a mass ratio $m_{\rm young}/m_{\rm old} = 0.12$ and a velocity dispersion of $\sigma = 30$ km~s$^{-1}$. At this mass ratio, the young population will dominate the light output for the younger ages, but will begin to blend with the older population at older ages, which we have limited to a 3 Gyr maximum. For the trials in the 1 Gyr to 3 Gyr range, some of the Monte Carlo trials did not converge very well, leading to volatility in the solutions. 

Other features in Figs. \ref{fig:Q8000} and \ref{fig:QADEV} are worth noting. Fig \ref{fig:Q8000} shows that Monte Carlo scatter is biased about 2\% toward older ages for both the CP-CP and non-CP-non-CP ``galaxy''-grid pairings (for ages for which all the Monte Carlo solutions seem reasonable). One also fails to see any difference between CP-CP and non-CP-non-CP inversions, pounding another nail into the coffin of expectation that switching to the CP model grid would automatically improve age discrimination. The template-mismatch case, however, shows a distinct bias toward older ages for the older component even though Fig. \ref{fig:QADEV} shows similar Monte Carlo scatter. Finally, Fig. \ref{fig:QADEV} shows an expected effect of the mass-to-light ratio wherein young populations are bright. At young ages the 12\% mass ratio is overcome by the brightness of the young subpopulation, but by the time that younger component ages to 3 Gyr, the 12\% mass ratio means its light is swamped by the 88\% dominant older 8 Gyr population. Therefore, the ADEVs are low for the young population at young ages, but climb, while the old-component ADEVs fall.

All in all, this modeling exercise showed itself to be inadequate, and yet pointed us toward eventual success. First, we must work to more effectively separate spectral effects due to the CP phenomenon from (1) continuum shape effects, and (2) other, more universal temperature-sensitive spectral features such as Balmer line strengths. Using least-squares regressions or, as we have done in this work, analyzing a hopeful sm\"org\aa sbord of indicators, are insufficiently targeted to uncover the CP signals in galaxies.

\section{Local Group Galaxies}

Spectral features from CP stars appear in many of the younger bins of averaged spectra from SDSS \citep{2015A&A...580L...5W}. Doubtless, however, many of the galaxies included in those grand averages did not contain the population ages where CP stars are prominent, and would not contribute to the signal. Individual galaxies with high S/N noise spectra present themselves as a fruitful avenue of investigation. To this end, we present observations of the near-nuclear regions of local group galaxies NGC~205 (M~110), NGC~221 (M~32), and NGC~224 (M~31).

\subsection{Observations}

\begin{figure}
	\includegraphics[width=\columnwidth]{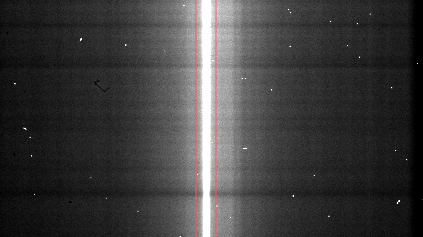} 
    \caption{A portion of one 2-d spectrum exposure of NGC~205 in the green. The color stretch is black for the sky level and white for maximum. The dispersion axis is vertical, with red upwards and the spatial axis is horizontal. The extraction aperture is shown by vertical red lines. In the case of NGC 205, the entire nuclear star cluster is contained within the aperture.}
    \label{fig:2d}
\end{figure}

Spectra were obtained with the Mark III spectrograph at the 2.4 meter telescope of MDM Observatory in 1993 September and 1994 October using a blue-sensitive CCD detector ("Charlotte"). The long slit was 2.36$\arcsec$ by 5.4$\arcmin$. Sky for NGC 224 was taken with a different pointing, but sky for NGC 205 and NGC 221 was extracted from the slit ends. CCD reductions were accomplished with IRAF \citep{1993ASPC...52..173T} to the point of sky-subtracted, wavelength-calibrated spectra. A correction to approximate spectrophotometricity was obtained by comparing our NGC 224 spectrum to that of \cite{1981ApJ...249...48G} and fitting the ratio with a low-order polynomial. The final spectra run from approximately 3800\AA\ to 6100\AA\ sampled at approximately 2.3 \AA\ per pixel. Spectra were extracted from the central $\pm$10 pixels (see Fig. \ref{fig:2d}) for an aperture full-width of 15\farcs6, or 56.9 pc at the distance of M~31 \citep{2012ApJ...745..156R}

The spectrograph focus was wavelength-dependent, degrading at both blue and red ends. With a wavelength-dependent convolution, we smoothed the spectra to a 200 km s$^{-1}$ standard. Following \cite{2014ApJ...783...20W}, we measured 80 Lick-style indices from the spectra. This allows us to exploit their index-based inversion program, COMPFIT.

\begin{figure}
	\includegraphics[width=\columnwidth]{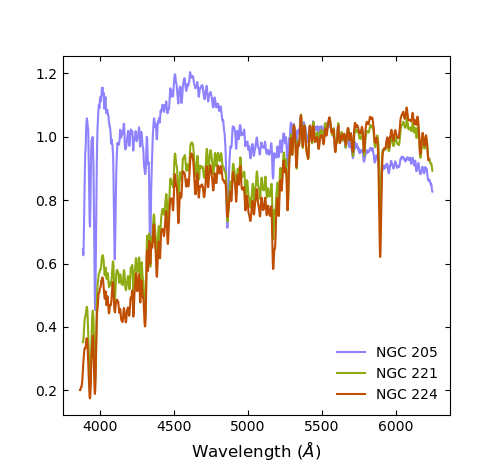} 
    \caption{Spectra of the three local group galaxies at instrumental resolution. Photon statistical plus CCD noise errors per pixel are inside the line widths. Spectra are normalized between $5300 < \lambda < 5800 $ \AA\ to show the color change.}
    \label{fig:3spectra}
\end{figure}

\subsection{Age structures}

We use two tools to comment on the star formation histories of these local group galaxies. One is to inspect for CP features and draw inferences from that. The other is to use a variant of our inversion program COMPFIT that we call AGEFIT. COMPFIT inverts the updated models as described in $\S$2 but includes an abundance distribution function (ADF) rather than a single metallicity \citep{2014ApJ...783...20W,2022MNRAS.tmp..287W}. Element-dependent isochrone shifts are included, and the CP stars as described here are included. The index set, however, is the one we have been using for years and does not contain any indices specialized to isolate CP features. The comparisons were carried out at a common velocity dispersion for models and galaxies of 200 km s$^{-1}$. The AGEFIT variant allows for three single-burst ages with a common abundance distribution function applied to all. If abundance ratios or IMF parameters are allowed to vary (we suppressed those variables, here) then they would apply to all three star formation bursts.

Previous stellar populations work on the nuclear regions of NGC~224 indicate dominance by an old, metal-rich system \citep{2000AJ....119.1645T}. The main sequence turnoff masses for such old populations are cooler than $\sim$6800 K where the cool end of the Am star sequence lies, so the CP method cannot be applied to this system. We see no evidence for an Am signature, and the AGEFIT results emphatically confirm an ancient, metal-rich population for the near-nuclear regions of this spiral galaxy.

NGC 221, on the other hand, has long been suspected of harboring an intermediate-age population \cite{1980ApJ...236..430O,2000AJ....119.1645T,2004AJ....128.2826W} in the 3 - 5 Gyr range. This "light-weighted mean age" could mean many things if one posits multiple populations. One possibility is that an underlying old population dominates in mass, but is overlain by a modest mass of younger stars. If that parcel of younger stars is young enough, it might enter the regime where CP stars become visible. O'Connell put the main sequence turnoff at F4, where the stars have convective envelopes and would not exhibit CP signatures. But, just a few hundred degrees hotter, F3 stars have gone radiative and have a $\sim$10\% fraction of Am stars \citep{1996Ap&SS.237...77S}. 

Our eyes, once again, fail to detect any Am-like spectral variations. AGEFIT results indicate that 90\% of the mass lies in a population 4.7 Gyr old, with 10\% in an ancient population. The mass ratios should not, at this stage of analysis, be considered secure. The third age component recommended by AGEFIT is about 1\% by mass at 950 Myr. The M/L ratio for young populations inflates this to several percent of the light output, increasing toward the UV. The metallicity that the program prefers has an ADF peak of [M/H]$_{\rm peak} =$ 0.24, or mass average [M/H]$_{\rm avg} =$ 0.05.

\begin{figure}
	\includegraphics[width=\columnwidth]{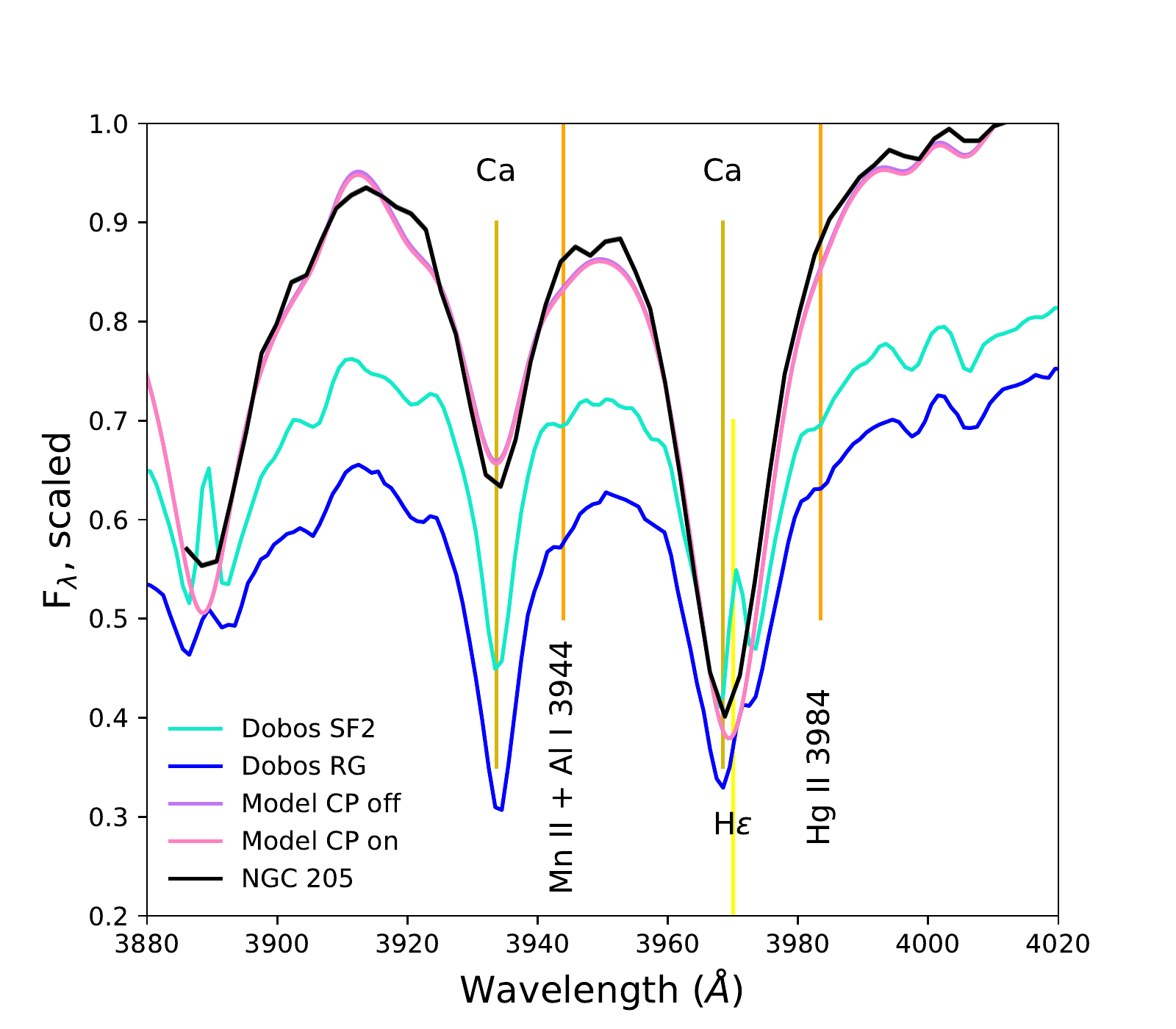}  
    \caption{Spectra in the blue for NGC~205 (black), SDSS averages (aqua and blue), and composite age models without the CP stars (purple) and with (pink) are compared. The model is 3\% by mass 70 Myr and the remainder at 10 Gyr, at solar composition. Neither model nor NGC~205 shows the Mn or Hg features that the SDSS averages do. Other points of interest are discussed in the text.}
    \label{fig:n205}
\end{figure}

NGC 205 has a blue nucleus and strong Balmer absorption. Its nucleus is known to contain clusters of B-type stars \citep{2009A&A...502L...9M}. This galaxy has a high probability of containing CP stars of the categories we investigate and thus presents itself as a detectability test. This detectability test runs into immediate problems, as shown in Fig. \ref{fig:n205}. We find no HgMn absorptions at all, the HgMn being the strongest and most emphatic of our CP categories in terms of spectral impact. Averaged spectra from SDSS, on the other hand, do (Fig. \ref{fig:n205}). The lack of HgMn features for NGC~205 rules out a strong contribution from turnoff stars in the 10700 K to 14000 K range, seeming to exclude ages between 100 Myr and 250 Myr.

NGC~205 does not show helium features, or a feature at 3920\AA\ probably due to C II and seen at spectral types B1 -- B3 \citep{2009ssc..book.....G}. This excludes the significant presence of O through B3 stars, and gives a hot limit to the near-nuclear population of $\approx$16000 K, corresponding to a turnoff mass of $m < 5.5$ M$_\odot$ and a young age of limit of 60 Myr. Given also its lack of nebular emission, it seems that NGC 205 is past its most recent phase of active star formation by at least 60 Myr.

\begin{figure}
	\includegraphics[width=\columnwidth]{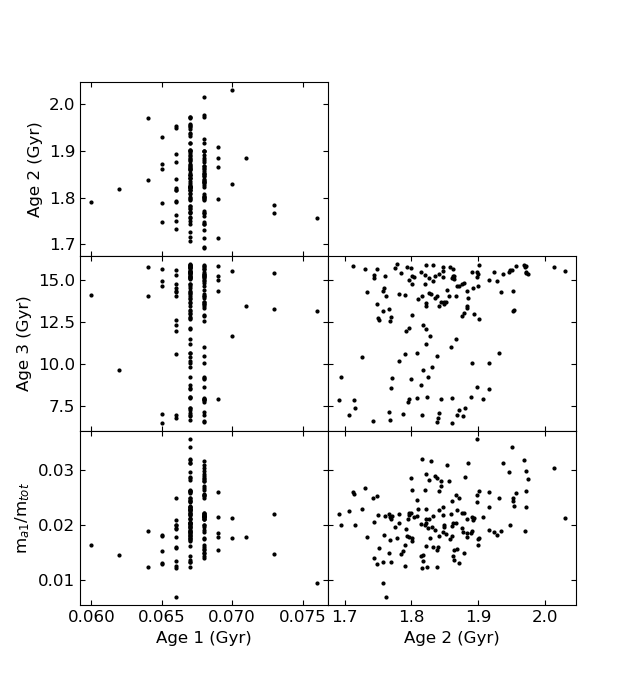} 
    \caption{Example parameter inversions for 200 Monte Carlo trials to get a sense of stability and parameter covariance. The target is NGC~205, compared with a normal-width ADF that peaks at solar metallicity. Some wild points are not shown. Age 1, Age 2, and Age 3 are the "young," 
    "intermediate," and "old" ages returned by AGEFIT, and $m_{a1}$ is the mass fraction of the young population. A 10-Myr discretization effect is seen for Age 1.}
    \label{fig:n205five}
\end{figure}

The above considerations seem to limit the age range allowed for the young component of NGC 205 to lie between 60 Myr and 100 Myr. It is remarkable, therefore, that AGEFIT threads the needle and falls within the window set by CP absorption features. Its preferred young age component is 68$\pm$2 Myr old at a mass fraction of 2$\pm$1\%, as illustrated in Fig. \ref{fig:n205five}. This population is so young, however, that it dominates the light despite its low relative mass. AGEFIT also finds that the strongest component in terms of mass is 1.85$\pm 0.1$ Gyr of age, with a weak ancient component. These older components, however, tended to drift and sometimes attempted to merge. The light-dominant 68 Myr component is well-constrained, but the exact mix of older components is not. 

Due to the list of things the models do not consider, the most significant omissions being non-bursty star formation and disconnected metallicities for each burst, these numerical results are unlikely to hold when measured against future attempts. 

\section{Conclusion}

This paper describes a modeling effort to include common types of chemically peculiar (CP) stars into integrated light models for stellar populations. Helium-weak, HgMn, Ap, and Am stars were included. These CP stars are on or near the main-sequence. Average surface abundance patterns were taken from the literature for each CP type, and synthetic stellar spectra were computed. Combined with observed estimates for the number fraction at each spectral type, the altered spectra were folded into existing isochrone-based integrated light models.

The synthetic spectra were compared with observed stars to confirm their adequacy. The model spectra for simple stellar populations (SSPs) were analyzed with an eye for extracting age information from the rise and fall of CP-caused absorption features. The spectral signatures are generally quite subtle, and even those subtle signatures attenuate quickly when high velocity dispersions are applied to the spectra. The spectra are quite rich, however. Significant age information is encoded there. Tests of a trial parameter inversion code show that up to three age bursts can be recovered in favorable cases. However, our first try to isolate the information coming from CP stars and consider it separately did not succeed, in the sense that the inversion program also utilized a fair amount of continuum shape information to arrive at its solutions. More work needs to be done to optimize CP stars as age indicators.

We did show, however, that template-mismatch is crucial when subtleties such as multi-burst solutions are sought. Even with ideal templates, neglecting the CP component led to $\sim$15\% age shifts in the older component of the 2-burst solutions.

Near-nuclear spectra for local group galaxies NGC~205, NGC~221, and NGC~224 are published here for the first time, though they were obtained some years ago. Spectral indices measured from these galaxies and fit with a three-age inversion program that included the CP-sensitive model improvements. CP-sensitive absorption features were not found in the spectra of any of the three galaxies. In the case of NGC~205, this led to strict exclusion zones for the possible age mixture. The inversion program output was in accord and in a three-burst model for the star formation history of NGC~205 reports a 2$\pm$1\% (by mass) component at 68$\pm$2 Myr that dominates the light, a mass-dominant component at 1.85$\pm$0.1 Gyr, and some fraction of very old stars for which the exact age and mass fraction is not well-constrained. 

Overall, the CP technique for uncovering star formation histories from integrated light continues to show promise, and continues to present challenges. Another lesson from this exercise is that more ordinary but subtle spectral features such as those used for over a century for spectra classification can be very sensitive to temperature and also encode star formation history information. High-signal galaxy and star cluster observations combined with yet-more realistic synthetic spectra inside the integrated light models should yield excellent results in the future.

\section{Data availability}

Much of the data underlying this article are publicly available. Any additional data underlying this article will be shared upon reasonable request to the corresponding author.

\section*{Acknowledgements}    

This work is based on observations obtained at the MDM Observatory, operated by Dartmouth College, Columbia University, Ohio State University, Ohio University, and the University of Michigan. We would also like to thank R. L. Kurucz for his freely available line lists and codes and the referee for his or her insight and drawing our attention to the dominance of Al I $\lambda\lambda$3944 compared to Mn II $\lambda\lambda$3944 at solar abundance ratios.

\bibliographystyle{mnras}
\bibliography{cpM32.bib} 

\bsp	
\label{lastpage}
\end{document}